\documentclass{pasj00}
\SetRunningHead{Astronomical Society of Japan}{Automated emission-line detection software for Subaru/FMOS}

\Received{2013/05/30}
\Accepted{2014/09/28}

\usepackage{times}

\begin{document}

\title{FIELD: Automated emission-line detection software for Subaru/FMOS
  near-infrared spectroscopy} 

\author{Motonari \textsc{Tonegawa},\altaffilmark{1}
  Tomonori \textsc{Totani},\altaffilmark{1}
  Fumihide \textsc{Iwamuro},\altaffilmark{2}
  Masayuki \textsc{Akiyama},\altaffilmark{3}
  Gavin \textsc{Dalton},\altaffilmark{4,5}
  Karl \textsc{Glazebrook},\altaffilmark{6}
  Kouji \textsc{Ohta},\altaffilmark{2}
  Hiroyuki \textsc{Okada},\altaffilmark{1}
  and Kiyoto \textsc{Yabe}\altaffilmark{7}
%  \thanks{Last update: April 21, 2014}
}

\altaffiltext{1}{Department of Astronomy, Faculty of Science, The University of Tokyo, 7-3-1 Hongo, Bunkyo-ku, Tokyo 113-0033}
\email{tonegawa@astron.s.u-tokyo.ac.jp}
\altaffiltext{2}{Department of Astronomy, Faculty of Science, Kyoto University, 
  Sakyo-ku, Kyoto 606-8502}
\altaffiltext{3}{Astronomical Institute, Faculty of Science, Tohoku University, 6-3 Aramaki, Aoba-ku, Sendai, Miyagi 980-8578}
\altaffiltext{4}{Department of Physics, University of Oxford, Keble Road, Oxford, OX1 3RH, UK}
\altaffiltext{5}{STFC RALSpace, Harwell Oxford, OX11 0QX, UK}
\altaffiltext{6}{Centre for Astrophysics \& Supercomputing, Swinburne University of Technology, P.O. Box 218, Hawthorn, VIC 3122, Australia}
\altaffiltext{7}{National Astronomical Observatory of Japan, 2-21-1 Osawa, Mitaka, Tokyo 181-8588}


\maketitle

\begin{abstract}
We describe the development of automated emission line detection
software for the Fiber Multi-Object Spectrograph (FMOS), which is a
near-infrared spectrograph fed by $400$ fibers from the $0.2$ deg$^2$
prime focus field of view of the Subaru Telescope. The software, FIELD
(FMOS software for Image-based Emission Line Detection), is developed
and tested mainly for the FastSound survey, which is
targeting H$\alpha$ emitting galaxies at $z \sim 1.3$ to measure the
redshift space distortion as a test of general relativity beyond $z
\sim 1$. The basic algorithm is to calculate the line signal-to-noise
ratio ($S/N$) along the wavelength direction, given by a 2-D
convolution of the spectral image and a detection kernel representing
a typical emission line profile.  A unique feature of FMOS is its use
of OH airglow suppression masks, requiring the use of flat-field images to
suppress noise around the mask regions. Bad pixels on the detectors
and pixels affected by cosmic-rays are efficiently removed by using
the information obtained from the FMOS analysis pipeline. We limit the
range of acceptable line-shape parameters for the detected candidates
to further improve the reliability of line detection.  The final
performance of line detection is tested using a subset of the
FastSound data; the false detection rate of spurious objects is
examined by using inverted frames obtained by exchanging object and
sky frames.  The false detection rate is $< 1$\% at $S/N > 5$,
allowing an efficient and objective emission line search for FMOS data
at the line flux level of $\gtrsim 1.0 \times 10^{-16}$[erg/cm$^2$/s].
\end{abstract}

\section{Introduction}

FMOS (the Fiber Multi-Object Spectrograph) is a near-infrared (NIR)
fiber-fed spectrograph for the Subaru Telescope, which is capable of
collecting nearly 400 spectra in a $0.2$ deg$^2$ field-of-view
(\cite{Kimura}) available at the prime-focus of the telescope.
The 400 fibers are divided into two groups of 200 fibers and
  connected to the two spectrographs, IRS1 (Infra-Red Spectrograph)
  and IRS2, for spectroscopy.
A feature of FMOS is the suppression
of bright OH-airglow emission lines using OH mask mirrors, which
reduce the OH-airglow emission lines by more than $90\%$
(\cite{Iwamuro2001}).  Since these airglow lines are the largest noise
source in the NIR region, the OH suppression allows FMOS to perform
NIR spectroscopic observations with a substantially reduced background
level.

NIR spectroscopy is a useful probe for a variety of topics in
astronomy.  The detection of emission lines (e.g., H$\alpha$\,$\lambda 6563$,
[NII]\,$\lambda \lambda6548,6483$, [OIII]\,$\lambda
\lambda4959,5007$)) in galaxies provide us with information
on diverse galaxy properties including the redshift, star formation rate, metallicity
and ionisation state.  A cosmological redshift survey, FastSound
\footnote{http://www.kusastro.kyoto-u.ac.jp/Fastsound/}, is ongoing, using FMOS to
measure $\sim 4,000$ redshifts of star-forming galaxies at $z\sim1.3$
from H$\alpha$ lines over a total area of $\sim 30\;{\rm deg^2}$
(\cite{Tonegawa}).  The main science goal is the measurement of redshift space
distortion (RSD) for the first time in this redshift range, to measure
the structure growth rate as a test of general relativity on
cosmological scales (see, e.g., \cite{Hawkins}; \cite{Guzzo};
\cite{Blake}; \cite{Samushia}; \cite{Reid}; \cite{Beutler};
\cite{Torre} for recent galaxy redshift surveys whose scientific
targets include RSD).

A key for studies treating many emission line galaxies in a large data
set is to automatically search and detect emission lines.  In the case
of FMOS, residual OH airglow lines just outside the OH mask regions and
cosmic rays often show spectral shapes similar to those of real
emission lines.  If such spurious line detections are included into a
statistical sample of emission lines, they will cause a systematic
error in, e.g., measurements of galaxy clustering power spectra.
Therefore, an efficient automatic line detection software with
minimized probability of false detection is highly desirable.

In this paper, we present the development of an emission line
detection software for FMOS spectroscopy, FIELD (FMOS software for
Image-based Emission Line Detection).  Two-dimensional spectral images
for each fiber are used in this algorithm in order to detect faint
emission line features, while efficiently filtering out unwanted
spurious detections.  We examine the performance of this routine by
applying it to a part of the FastSound project data. We test various
parameters about shapes of emission lines in the 2D spectral images,
to cut spurious objects efficiently. The contamination fraction of
spurious objects is estimated by inverted 2D spectral images, which
are obtained by exchanging object frames and sky frames.

In Section \ref{Sample}, we describe the data used in this work.  In
Section \ref{Method}, we present the algorithm for detecting
emission line candidates and describe key features that effectively
reduce false detections caused by residual OH emission lines and
cosmic rays.  The results are presented in Section \ref{Results} and we
summarize this work in Section \ref{conclusion}.

\section{The Sample}\label{Sample}

\subsection{FMOS Observation}

The targets for FMOS spectroscopy were selected using photometric
redshift and H$\alpha$ flux estimates using $u'g'r'i'z'$ data of
Canada-France-Hawaii Telescope Legacy Survey (CFHTLS) Wide
(\cite{Gwyn}; \cite{Tonegawa}).  Observations are done with the
normal-beam-switch (NBS) mode of FMOS, in which the object frame is
taken using nearly all $400$ fibers for targets, followed by an offset-sky
frame with the same fiber configuration.  In FastSound, the
exposure time for one frame is $900$ s, and two frames (i.e., total
exposure of $30$ min) are taken both for the object and sky frames. We use the
high-resolution (HR) mode with spectral resolution of
$R\sim2200$ and the wavelength range of $1.44$--$1.66\;\mu$m.
The detector size is 2k$\times$2k, and the pixel scale is $1.1 \ {\rm
  [\AA/pix]}$.

The primary FMOS data set used in this paper are eight FMOS
field-of-views (FoVs) observed during March 28--29 and April 1--2 of 2012,
as part of the FastSound Project.
The FastSound field IDs are W2\_041, 062, and 063 from the CFHTLS W2
field, and W3\_162, 163, 164, 177, and 192 from W3.  (See a
forthcoming FastSound paper Tonegawa \etal, in prep. for details of the FastSound
observations.)  FMOS fibers were allocated to $2,871$ galaxies in the
eight FMOS FoVs and the emission line detection rate is typically
$\sim 10\%$ as reported in Tonegawa et al. (2014).  The mean seeing is
$\sim 0.8''$ in the $R$ band. These eight FoVs are representative of
FastSound data in normal observing conditions. 

In rare cases (four out of $119$ FoVs in the FastSound project),
images taken by IRS2 spectrograph just after sunset show unusual dark patterns on
the edge of detector, likely because of the instability of detector
resulting from overexposures by dome-flat lamp or twilight.  These
patterns usually appear vertically in the left end ($1.44$--$1.45\;\mu$m) of the
image, but sometimes appear horizontally in the top or bottom end.
Though the fraction of such data is not large in the total
FastSound data set, line detection must be affected in these regions.
Hence we also examine the performance of our software in these four
FoVs in these bad conditions. These are W2\_042 (taken in 2013 Feb.), W3\_173
(2012 Apr.), W3\_206 and 221 (2012 Jun.).

\subsection{Basic Data Reduction}

The FMOS two-dimensional raw data are reduced with the standard data
reduction pipeline, {\it FIBRE-pac} (Iwamuro et al, 2012).  In FMOS
images, the $x$-axis is used for the wavelength direction, and $9$
pixels are assigned in the $y$-axis direction for each fiber.  The
profile in the $y$-axis is just the point spread function (PSF) of
light from a fiber.  First, sky subtraction is performed for each
frame. In order to trace the time-variance of sky background, the
linear combination of two sky frames is used for the subtraction.
After sky subtraction, the bias difference between the four detector quadrants, cross
talk, bad pixels, distortions, and residual OH lines are
corrected. The two images are then combined into the total exposure.
The wavelength calibration is applied by making a correlation between
$x$-axis and physical wavelength using images of a Th-Ar lamp.
\textit{FIBRE-pac} creates not only a science frame from object and
sky images, but also square-noise and bad pixel maps for each FoV
(Figure 1).

The square-noise map gives the square of noise level of each pixel, which is
measured using $200$ pixels around each pixel along the $y$-axis with a
$3\sigma$ clipping algorithm iterated ten times.  The bad pixel map
indicates the quality of each pixel by a value between between $0$ (bad)
and $1$ (normal).  Those pixels on the detector flagged as non-functioning by 
\textit{FIBRE-pac} and temporally prominent pixels appearing
in one exposure (by cosmic ray events in most cases) are rejected and
replaced in the science frames, by an interpolated value from
surrounding pixels (\cite{Iwamuro}). These pixels are stored as a bad
pixel map for each exposure of $15$ min.  Our exposure time is $30$ min,
and hence we have two bad pixel maps for each field-of-view.  When
these two maps are combined, true defects of the detector are presented as
the value $0$ on the bad pixel map because they appear in both of two
bad pixel maps, while cosmic rays have a value of $0.5$ because they
affect only one of the two science sub-frames.  The information from
the square-noise and bad pixel maps are utilized in the line detection
algorithm presented below.

As well as the normal science frame, we also create an ``inverted"
science frame by exchanging object and sky frame in the 
\textit{FIBRE-pac} procedures.  The line detection algorithm is
also applied to this inverted science frame.  Any candidates detected
in this inverted frames will not be genuine features, because
real emission lines are negative in the frame and absorption lines are
usually below the detection limit of FMOS for galaxies beyond $z \sim
1$. Therefore we can reduce the overall false detection rate by tuning the
software to minimize the detection rate in the inverted frames.

\section{The Line Detection Algorithm}\label{Method}

\begin{figure}
 \begin{center}
 \FigureFile(80mm,80mm){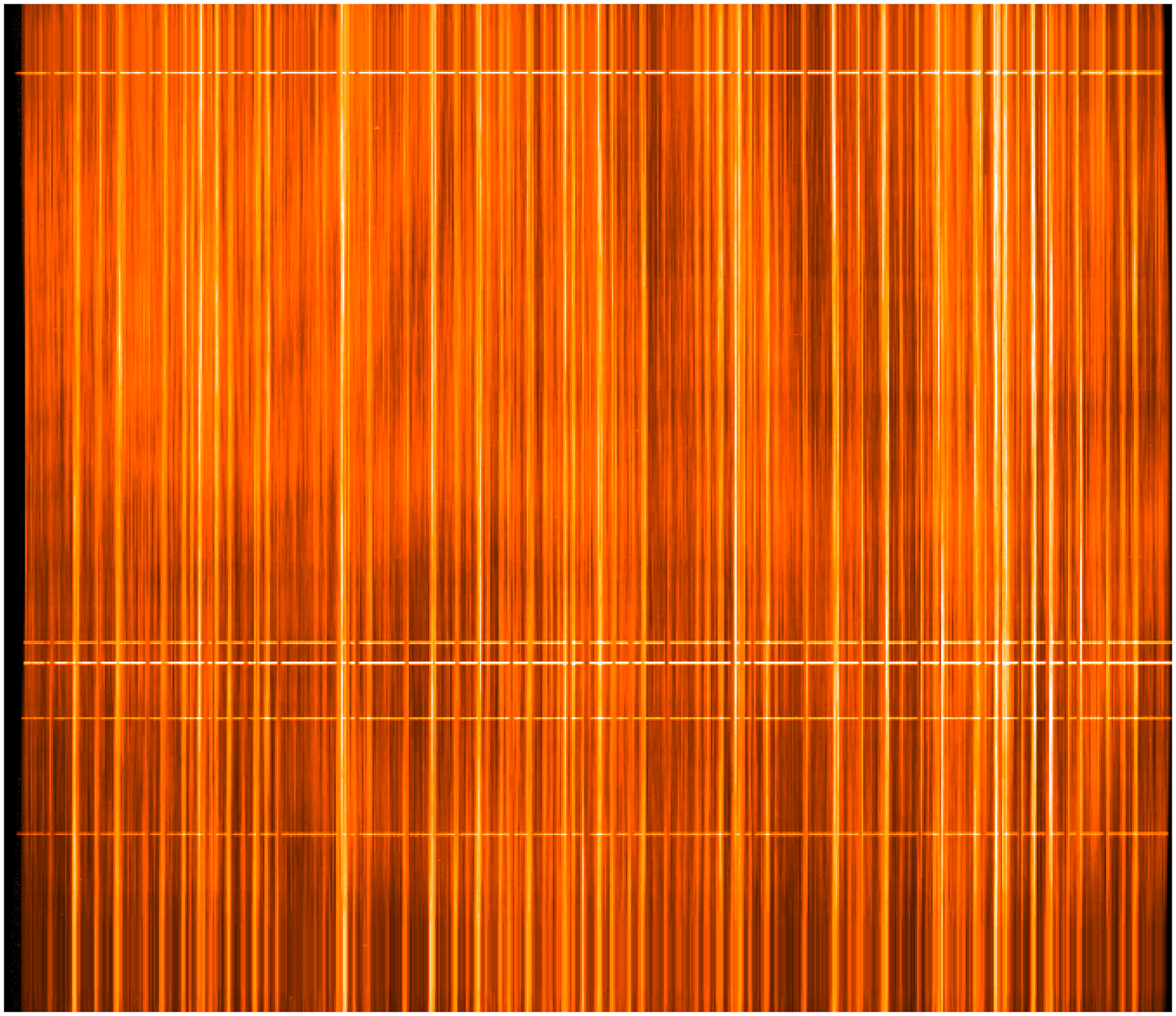}
 \FigureFile(80mm,80mm){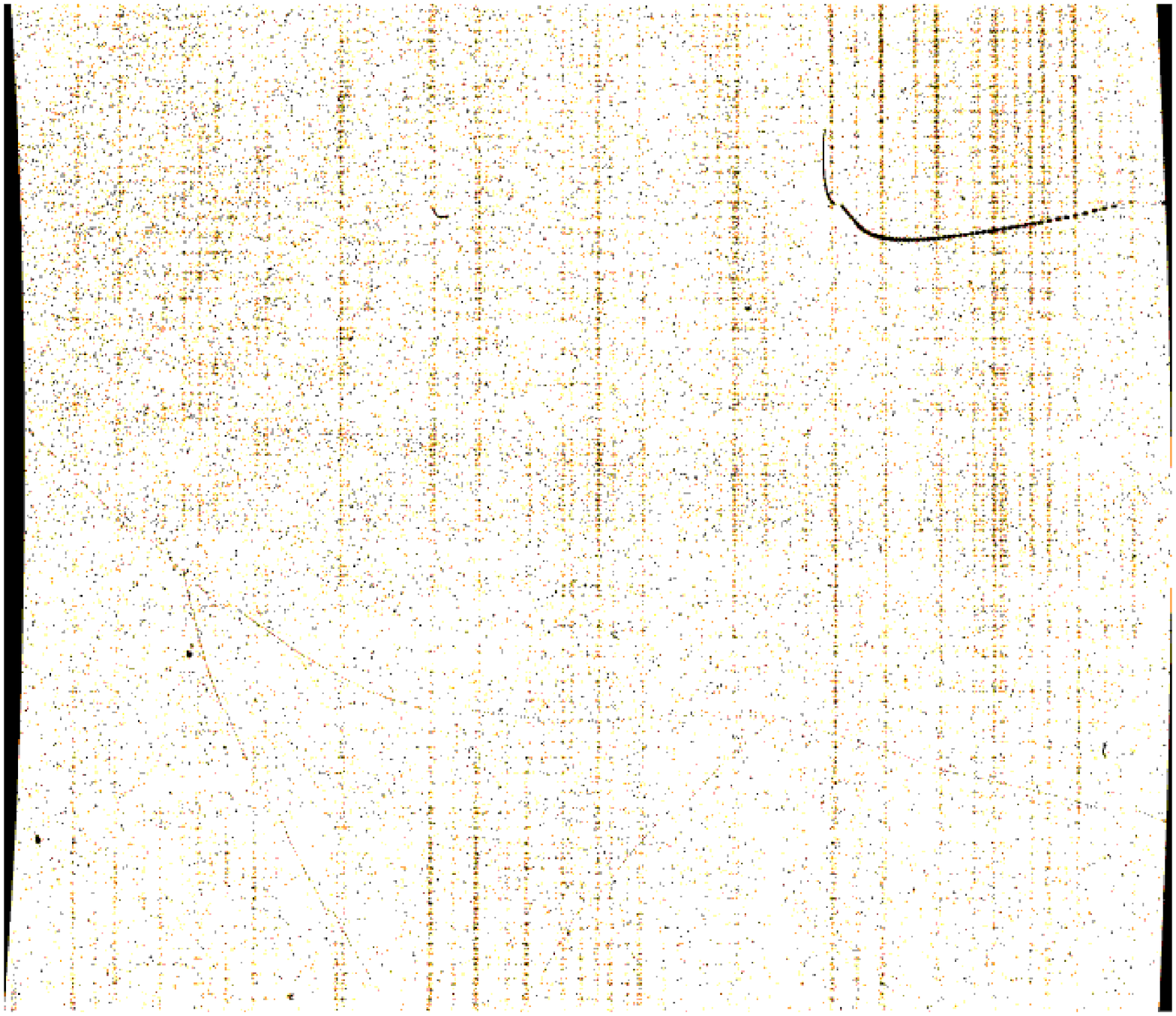}
 \end{center}
 \caption{({\it Top}) A square-noise level map drawn from an IRS2
   image of FMOS.  White regions represent higher noise level.
   Vertical high noise patterns trace the residual OH airglow emission
   lines.  (\textit{Bottom}) A bad pixel map drawn
   from an IRS2 image of FMOS.  White, orange, and black pixels
   represent the map values of $1$ (normal pixels), $0.5$ (mostly
   cosmic ray events), and $0$ (registered bad pixels), respectively.
   The curved feature in the top right quadrant is a scratch on the detector
   of the IRS2 spectrograph.}
 \label{figure:FMOSimage}
\end{figure}

Our goal is to automatically detect spectral features corresponding to
emission lines from the FMOS data with high reliability.  To achieve this,
we develop an emission line detection algorithm which is based on
a convolution of the FMOS 2-D spectral images weighted by a detection
kernel that is similar to the typical emission line profile
(see, e.g., \cite{Gilbank} for similar approaches).
Note that
we do not use the flux-calibrated 1-D spectra for emission line
detection because they do not retain the 2-D
shape of the PSF, which is useful to filter out false
detections.  The algorithm uses a flat-field image, square-noise map and bad
pixel map as well as the science frame (see \S\ref{Sample}).  The
science frame and square-noise map are both convolved with the kernel,
producing an effective signal-to-noise ($S/N$) ratio of a line
centered at each pixel along the wavelength direction.  The flat-field image
and bad pixel map are used to remove the OH mask regions
or bad pixels from the calculation of $S/N$, so that the false
detection rate is minimized.

\subsection{Image Processing Before Line Search}

An important feature of the FMOS spectra in the NIR is the hardware suppression of the
the OH airglow. Inclusion of the masked regions does not
improve the line detection efficiency, and the noise level is often
particularly high at the border of OH masks, which would have a
negative effect for efficient line detection. To remove such regions,
we utilize the dome-flat image. The dome-flat image is divided by the
detector-flat image to correct for different quantum efficiency
between pixels (\cite{Iwamuro}).  After normalization, the flat-field image
have values of $\sim1$ at normal pixels, while $\sim0$ at OH mask
regions.  We then make a new square-noise map by dividing the original
square-noise map by this flat image.  This operation increases the
noise level inside and around OH mask regions by a large factor,
making such regions ineffective in the line $S/N$ calculation (see the
next subsection).  This approach is better than, e.g., simply removing
OH mask regions by wavelength information, because the detailed performance
of the OH mask depends on the temperature of FMOS
instrument that changes with time.  However, the effect of noisy
regions around the border of OH masks cannot be completely removed by
this operation, and hence we further remove the pixels whose original
noise level (before dividing by the flat image) is higher than the
mean by more than $2\sigma$, from the line $S/N$ calculation.

In order to decrease the chance of detecting cosmic rays and detector
defects, we use of the bad pixel map (Figure
\ref{figure:FMOSimage}).  Since the bad pixel map has values of
$\sim1$ at good pixels, $\sim0$ at defects, and $\sim0.5$
at pixels hit by cosmic rays, we exclude the pixels having values
lower than $0.7$ in the bad pixel map from the line $S/N$ calculations.

Finally, the continuum component of galaxies may affect the detection
efficiency of emission lines. Therefore the continuum component of
each spectrum in the object frame was subtracted by applying fit1d
task of IRAF with a 5th-order Chebyshev polynomial along the
wavelength direction. This procedure was adopted separately for
  each pixel along the vertical direction perpendicular to
  wavelength.

\subsection{Candidate Selection by Line Signal-to-Noise}
\label{subsection:algorithm}

We define the detection kernel by a two-dimensional Gaussian that
imitates the typical shape of an emission line:
\begin{equation}\label{kernel}
g(x,y) \equiv \exp \left[-\frac{1}{2} \left(\frac{x^2}{\sigma_x^2} +
  \frac{y^2}{\sigma_y^2} \right) \right]
\end{equation}
where $\sigma_x$ and $\sigma_y$ are the typical dispersion of emission
lines along $x$-axis (corresponding to wavelength) and $y$-axis
(corresponding to the fiber aperture) on the detector.  Although
$\sigma_x$ and $\sigma_y$ are adjustable, we fix $\sigma_x=4.26$ [pix]
as the quadratic sum of the spectral resolution of FMOS ($6.375$ {\AA}
FWHM at $\lambda=1.53\;\mu$m in the HR mode) and the typical velocity
dispersion of galaxies ($175\;{\rm km/s}$ FWHM), and $\sigma_y=2.5$
[pix] as representative of the FMOS fiber profile on the
detector.
Because FMOS is a fiber-fed spectrograph,
$\sigma_y$ is just an instrumetal spread of light from fibers 
without spatial information of
target objects.

 For each pixel $(x_c,y_c)$, we define the
signal $S$ of a supposed line centered at the pixel as:
\begin{equation}\label{S}
S(x_c,y_c) = \alpha \sum_{(x_i,y_i)\in D}
g(x_i-x_c,y_i-y_c),
\end{equation}
which is an integration over pixels $(x_i, y_i)$ around $(x_c,y_c)$
confined within the region $D$. This region is an ellipse whose center
is $(x_c,y_c)$ and the radii along the major and minor axis are
$2\sigma_x=8.52$ [pix] and $2\sigma_y=5.0$ [pix].  The parameter
$\alpha$ is the best-fit flux normalization of this kernel to the
observed count $c_i$ in the pixels in $D$:
\begin{equation}\label{alpha}
\alpha = \frac{\sum_{(x_i,y_i)\in D}
  c_i\frac{g(x_i-x_c,y_i-y_c)}{n_i^2}}{
  \sum_{(x_i,y_i)\in D} \frac{g(x_i-x_c,y_i-y_c)^2}{n_i^2}} \ ,
\end{equation}
which can be derived by minimizing $\chi^2$, where
\begin{equation}
\chi^2 \equiv \sum_{(x_i,y_i)\in D} \frac{[c_i - \alpha
  g(x_i-x_c,y_i-y_c)]^2}{n_i^2} \ ,
\end{equation}
where $n_i$ is $1\sigma$ noise at $i$-th pixel from the noise-square
($n_i^2$) map. From the expression for $\alpha$, it can be understood
that the signal $S$ (Equation \ref{S}) is a convolution of the count
$c_i$ with the detection kernel, weighted by $1/n_i^2$.  We
then define the noise $N$ for the line; since $S$ is a linear
combination of $c_i$ whose noise is $n_i$, its statistical error can
be calculated as:
\begin{equation}\label{N}
N(x_c,y_c) = \frac{\sum_{(x_i,y_i)\in D}
g(x_i-x_c,y_i-y_c)}{
\sqrt{\sum_{(x_i,y_i)\in D} \frac{g(x_i-x_c,y_i-y_c)^2}{n_i^2}} } \ .
\end{equation}
Then the signal-to-noise ratio $S/N$ is simply calculated from $S$
and $N$.

The $S/N$ is calculated along the wavelength direction, with $y_c$
fixed at the central pixel ($y_c = 0$) among the 9 pixels for one
fiber.  When a certain pixel has a local peak of $S/N$ map, the pixel
is likely to be the center of an emission line feature.  Therefore, we
select emission line candidates by extracting pixels with $S/N$ values
higher than a threshold and locally greatest within the range of
$\pm$30 pixels from $x_c$ along the wavelength direction.  The latter
condition is introduced to avoid multiple detections of the same
line-like feature.  Although the use of $\pm$30 pixels as the
  minimum separation is a simplistic approach, it suffices for
  FastSound, because the closest lines (H$\alpha$\,$\lambda 6563$ and
  [NII]\,$\lambda \lambda6548,6483$) at $z\sim1.3$ are separeted by
  $\gtrsim40$ [pix] on the image.  More sophisticated methods to
  discriminate multiple lines, such as connected-pixel approach, will
  be examined in future work.  The selection threshold can be changed
by the user: if one increases the threshold, the number of false
detections decreases but the number of real lines would also be
reduced. The false detection rate and its dependence on the threshold
are discussed in \S\ref{Results}.

\subsection{Further Selection by Image Shape Parameters}\label{subsection:shape}

The image shape of each candidate line includes important information
that allows discrimination between real and false lines. 
We measure the following shape parameters:

\noindent (1) center along $y$-axis: 
\begin{equation}
 \overline{y}=\sum_{(x_i,y_i)\in D} y_i c_i \bigg/ \sum_{(x_i,y_i)\in D}c_i.
\end{equation}
(2) dispersion along $y$-axis:
\begin{equation}
 {\sigma_y}^2=\sum_{(x_i,y_i)\in D} (y_i-\overline{y})^2 c_i \bigg/ \sum_{(x_i,y_i)\in D}c_i.
\end{equation}
(3) fraction of positive pixels:
\begin{equation}
 f_p = \sum_{(x_i,y_i)\in D} H(c_i) \bigg/ \sum_{(x_i,y_i)\in D} 1 \ , 
\end{equation}
where $H(x)$ is a step function (1 for $x > 0$ and 0 otherwise).\\
(4) position angle of the major axis of an elliptical fit:
\begin{equation}
 \theta=\frac{1}{2}\arctan
 \left(2\frac{\overline{xy}}{\overline{x}^2-\overline{y}^2}\right)
\end{equation}
(5) axis ratio of the elliptical fit:
\begin{equation}
 \frac{b}{a} = \frac
{\frac{\overline{x^2}+\overline{y^2}}{2} -
   \sqrt{\left(\frac{\overline{x^2} -
       \overline{y^2}}{2}\right)^2+\overline{xy}^2} }
{\frac{\overline{x^2}+\overline{y^2}}{2} +
   \sqrt{\left(\frac{\overline{x^2}-\overline{y^2}}{2}\right)^2 +
     \overline{xy}^2}} \, 
\end{equation}
where $a$ and $b$ are semi-major and minor axes, respectively.\\
(6) signal-to-noise fluctuation per pixel:
\begin{equation}
 \sigma_{\rm SNR} = \left\langle \frac{c_i^2}{n_i^2} 
\right\rangle - \left\langle \frac{c_i}{n_i} \right\rangle^2 \ ,
\end{equation}
where angle brackets denote the mean over the region $D$. 

We applied our algorithm to an ``inverted" science frame, obtained by
running the FMOS image reduction pipeline with the object and sky
frames exchanged, as well as the ``normal" science frame.  All the
candidates in the inverted frames must be spurious, and their
statistical nature should be the same as the spurious features in the
normal frames, because the analysis procedures are exactly the same
except for swapping the object/sky frames. For example,
  residual OH emission lines in the science frame can be positive or
  negative at the same probability depending on the observing
  conditions, and cosmic rays fall randomly both on the object and sky
  frames.  Note that an emission line galaxy can accidentally fall in
a fiber during the sky exposures, but we did not take these cases into
account, because the possibility should be small.

\begin{figure*}
   \begin{center}
      \FigureFile(75mm,75mm){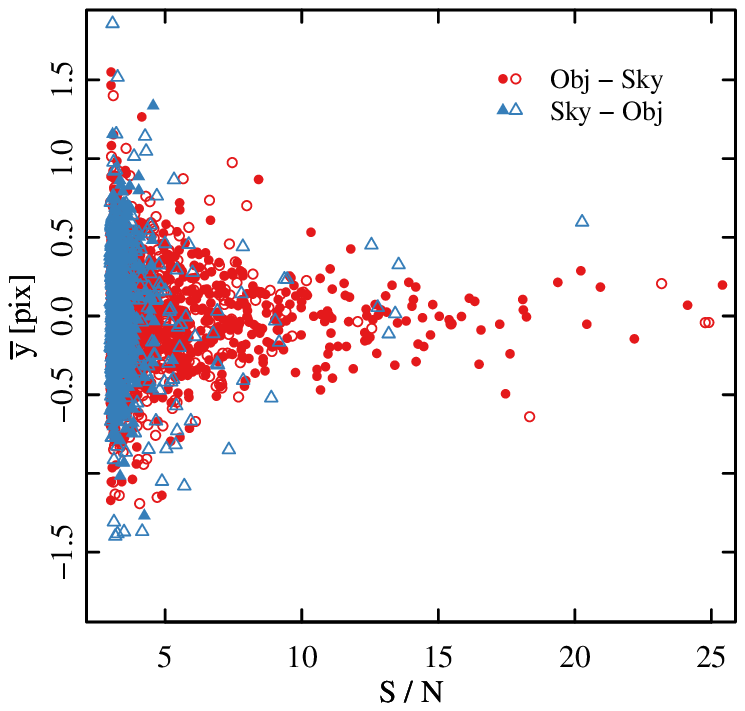}
      \FigureFile(75mm,75mm){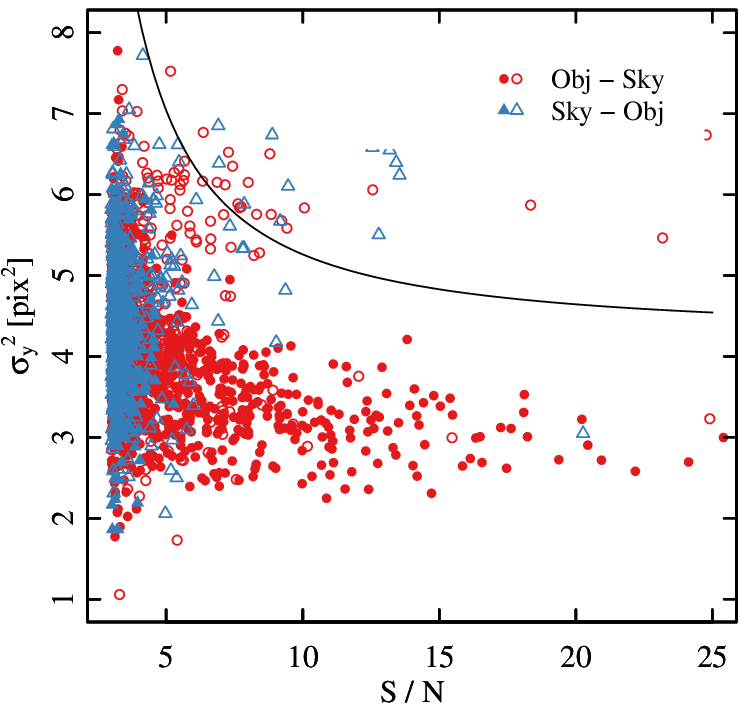}
      \FigureFile(75mm,75mm){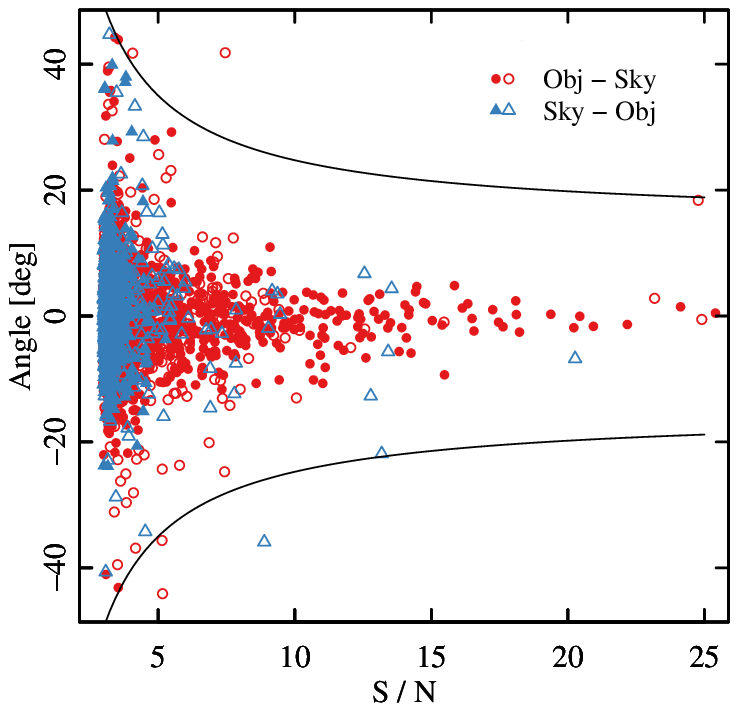}
      \FigureFile(75mm,75mm){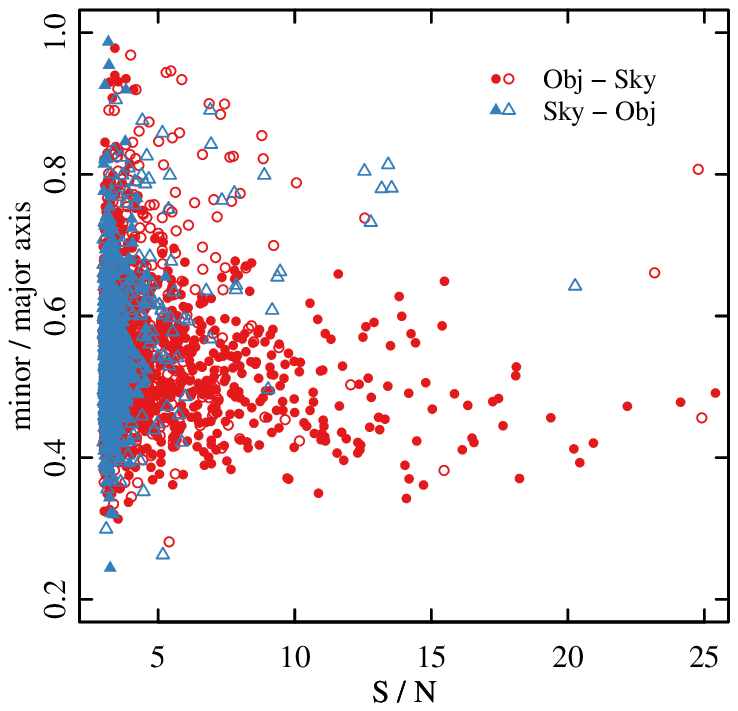}
      \FigureFile(75mm,75mm){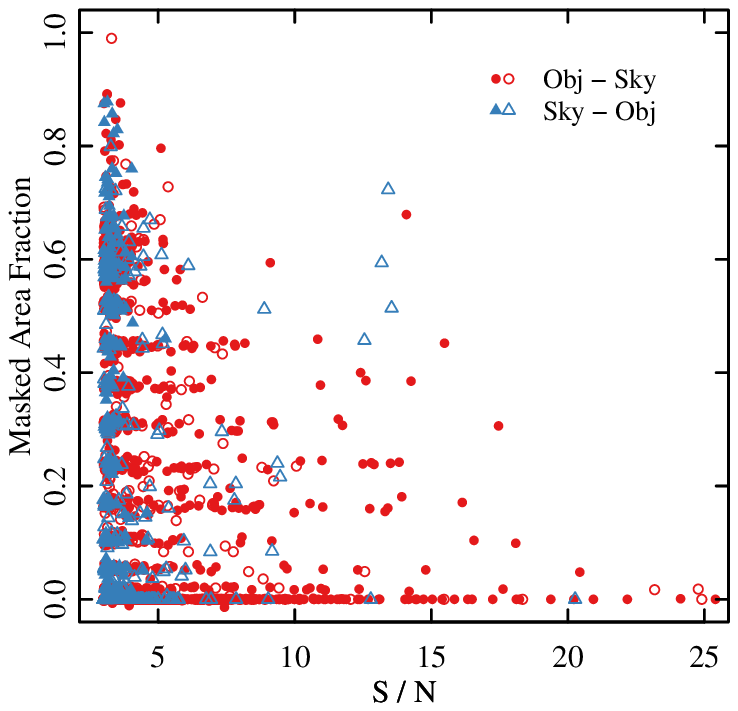}
      \FigureFile(75mm,75mm){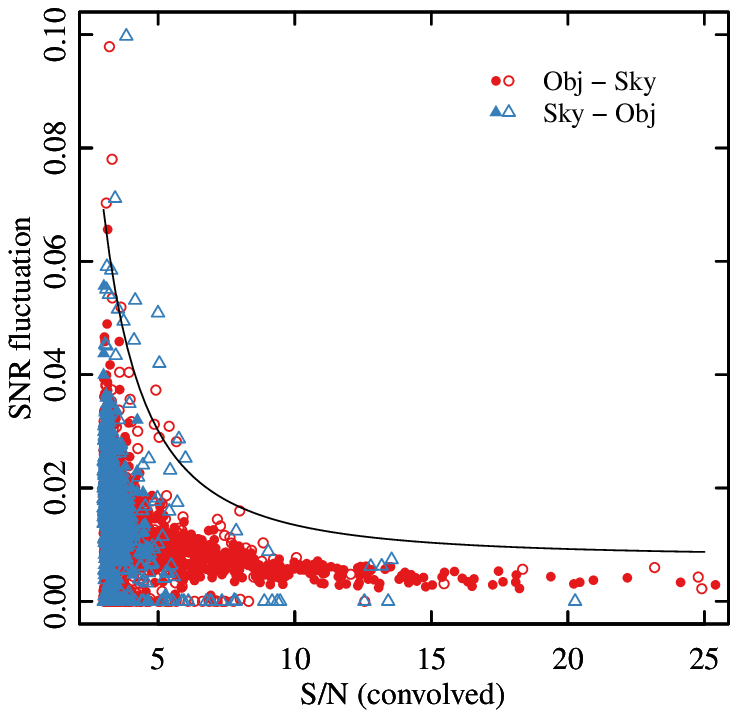}
   \end{center}
   \caption{The six shape parameters as a function of $S/N$, for the
     normal (red circles) and inverted (blue triangles) frames.  Filled symbols are used for line candidates
     in the eight FMOS FoVs taken in normal conditions and open symbols
     for the four FoVs taken in the bad condition (see \S \ref{Sample}).  The
     solid curves show the parameter cuts adopted to remove
     false-detections. In the top-left panel, the zero-point of $y$ is
     set at the midst of the nine pixels for each fiber. }
\label{figure:param}
\end{figure*}

Figure \ref{figure:param} shows the distribution of
the shape parameters defined above, as a function of $S/N$, both for
the normal and inverted frames.

\begin{figure*}
   \begin{center}
   \FigureFile(80mm,66mm){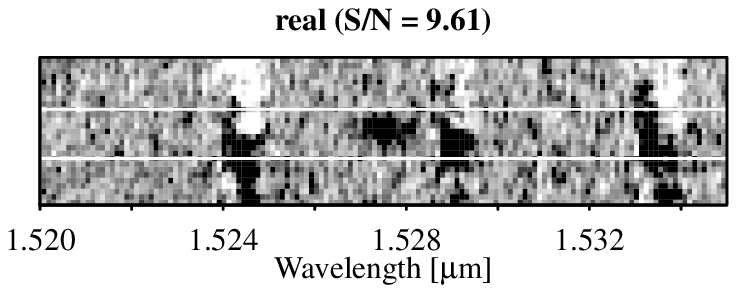}
   \FigureFile(80mm,66mm){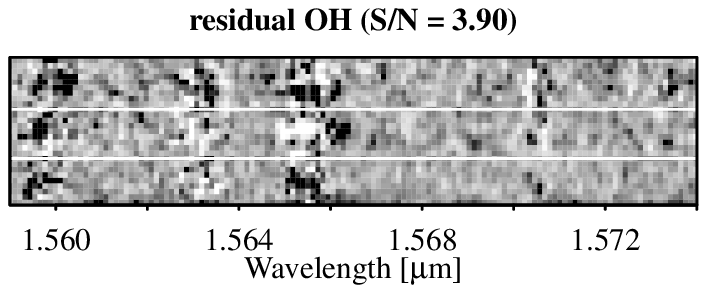}
   \FigureFile(80mm,66mm){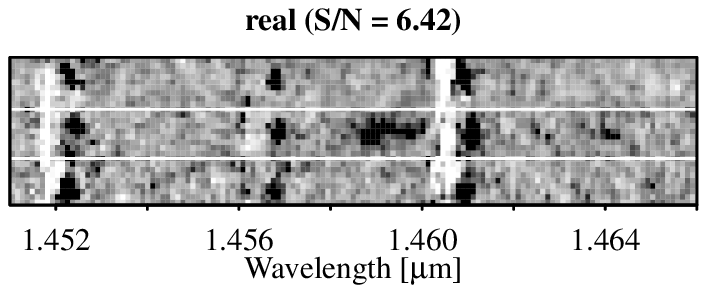}
   \FigureFile(80mm,66mm){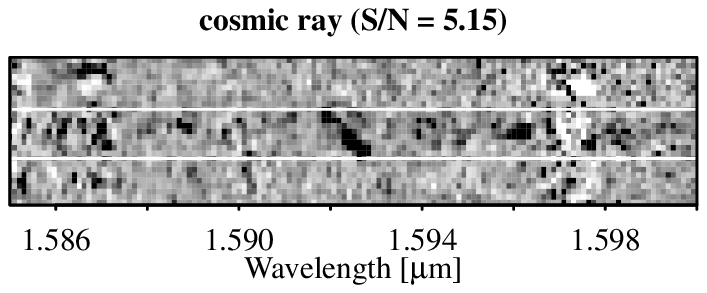}
   \FigureFile(80mm,66mm){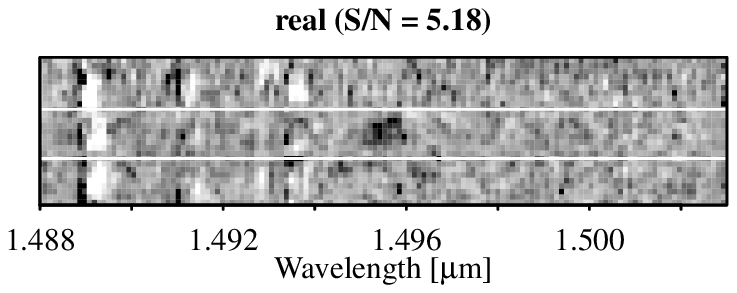}
   \FigureFile(80mm,66mm){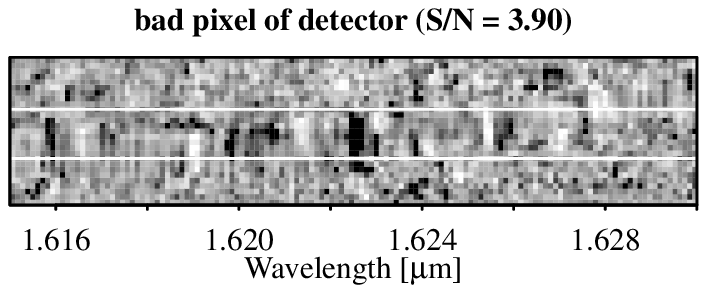}
   \FigureFile(80mm,66mm){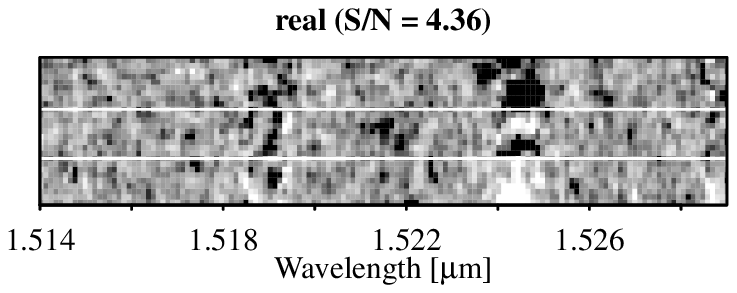}
   \FigureFile(80mm,66mm){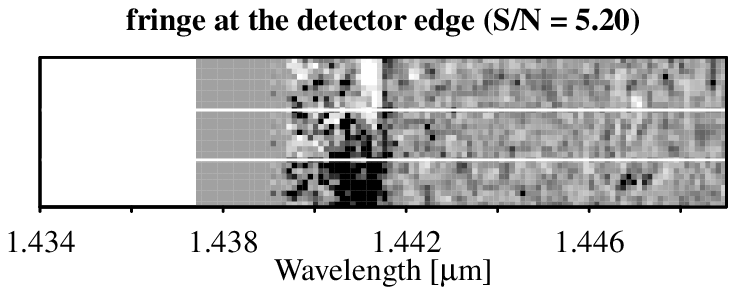}
   \end{center}
   \caption{Examples of 2-D images of eight emission line candidates
     detected by the software.  The white horizontal lines separate
     different fibers, and two neighboring fibers are also shown.
     The left four panels show those judged to be real by eye
     examination, in a range of $S/N = $ 4.36--9.61.  The right four
     panels are examples of false detections remaining even after
     the shape parameter cut. The causes of these false detections are
     indicated on the top of each panel.  Vertical features appearing
     in some panels are residual OH lines, except for the bottom-right
     panel that shows the case of unusual dark pattern at the edge of the detector
     in the bad condition (see \S\ref{Sample}).}
   \label{figure:sample}
\end{figure*}

It is found that in some regions (especially for the data in the 4
FoVs at the bad condition in the plots of $\sigma_{y}^2$, $\theta$,
$b/a$, and $\sigma_{\rm SNR}$ plots) the number of inverted-frame
objects is relatively large compared with the normal-frame objects,
and the numbers in normal/inverted frames are similar.  False lines
are expected to be dominant over real lines in such regions, and hence
we introduced an event cut as shown by the solid curves in the
figure. These conditions are expressed as:
\begin{eqnarray}
\sigma_y &>& 4.210 + 6.868 \left( \frac{S}{N} \right)^{-1}
 + 36.413 \left( \frac{S}{N} \right)^{-2} \\
|\theta {\rm \ [deg]}| &>& 15.00 + 94.79 \left( \frac{S}{N} \right)^{-1}
 + 26.30 \left( \frac{S}{N} \right)^{-2} \\
\sigma_{\rm SNR} &>& 7.701 \times 10^{-3} 
+ 3.609 \times 10^{-3} \left( \frac{S}{N} \right)^{-1} \nonumber \\
&& + 0.5416 \left( \frac{S}{N} \right)^{-2} \ .
\end{eqnarray}
It should be noted that the region of many inverted-frame objects is
also found in the plot of minor/major axis ratio. We did not include a
condition on this quantity, because we found that almost all of the
false objects in this region are effectively removed by the other three
conditions.  We also examined the spectral images of the rejected
events by eye, and confirmed that false detections do indeed dominate.
These events mainly arise from unusual dark patterns appearing in the bad-condition
four FoVs, when the detector was in an unstable state.  Examples of
emission line candidates detected by the software are shown in Figure
\ref{figure:sample}.

\section{Performance}\label{Results}

\begin{table}
\footnotesize
\begin{center}
  \caption{Detected line candidate statistics for the eight FoVs under
    normal condition.\footnotemark[*]
    }
\begin{tabular}{cccc}
\hline
\hline
  & Obj.-Sky (1) & Sky-Obj. (2) & Contamination (2)$/$(1)\\
\hline
$S/N>5.0$ & $325$ & $1$ & $0.3\%$ \\
$S/N>4.5$ & $381$ & $7$ & $1.8\%$ \\
$S/N>4.0$ & $480$ & $25$ & $5.2\%$ \\
\hline
\hline
\end{tabular}
\label{table:linestat}
\end{center}
\footnotemark[*]{Statistics for both the normal and inverted images are presented.
The contamination rate is estimated from the statistics of the inverted images.}
\end{table}

In Table \ref{table:linestat}, the number of emission line candidates
in the eight FoVs under normal condition is shown for three different
thresholds of $S/N$, both for the normal and inverted frames.  There
is only one false-detection in the inverted frame above $S/N=5.0$,
indicating that the false detection rate should be less than 1\% at
$S/N > 5$. We inspected the seven detections in the inverted frame at
$S/N > 4.5$ by eye, and found that five were close to OH mask region,
one at the edge of the detector, and the other one due to 
a cluster of hot pixels that shows a collective instability
because of high dark current. Although the line
detection algorithm suppresses the contribution to $S/N$ from the OH
mask region, strong residual OH emission often remains because of
unstable OH airglow in the NIR, resulting in a small fraction of false
detections.  Since the events at the border of OH masks are the major
source of the false detections, more strict cuts in such regions may
further improve the fraction of real lines.

\begin{figure*}
   \begin{center}
      \FigureFile(80mm,80mm){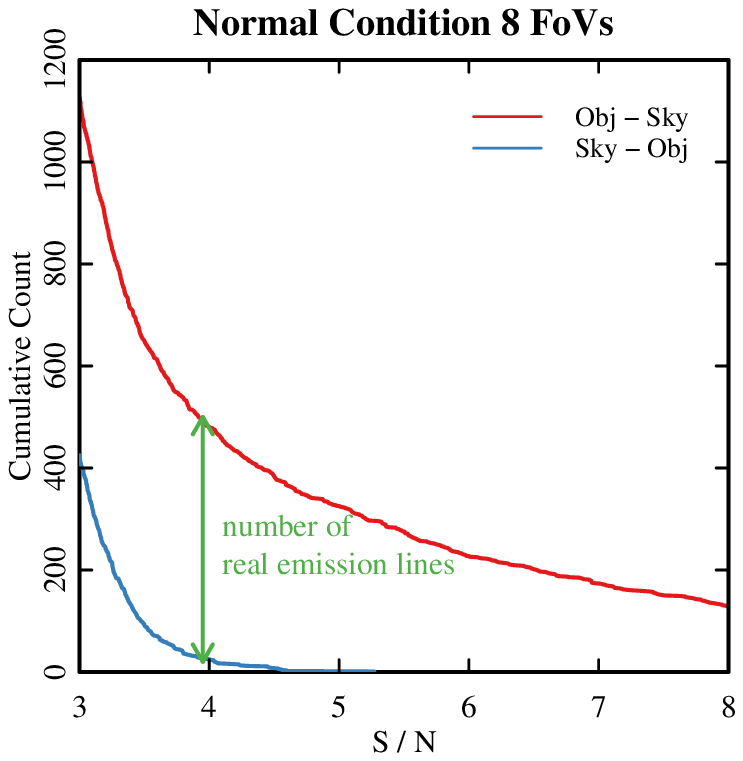}
      \FigureFile(80mm,80mm){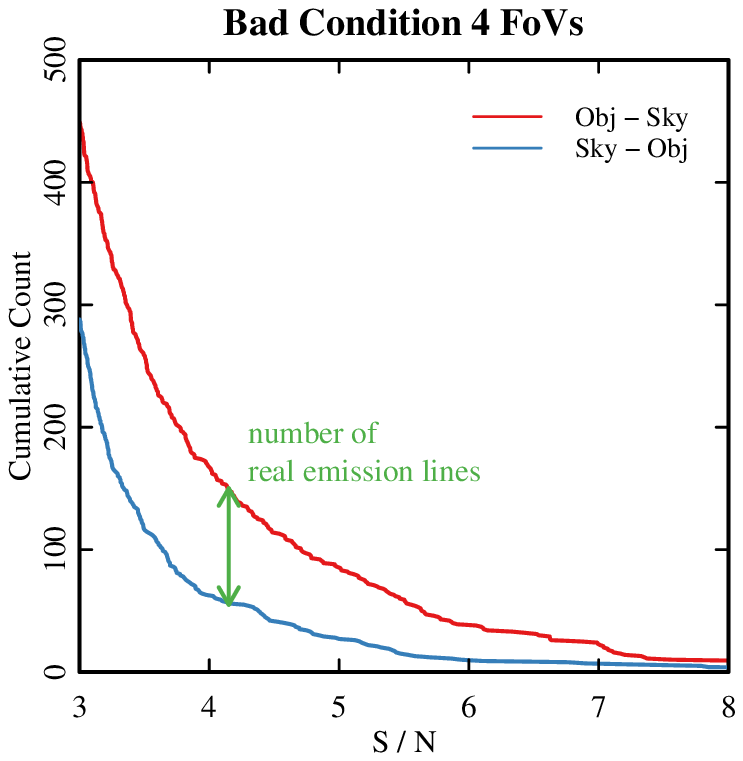}
   \end{center}
   \caption{The cumulative counts of emission line candidates
     above a given value of $S/N$ detected in the eight FoVs of normal
     condition (left) and the four FoVs of bad condition (right).  The
     red line shows the number of candidates detected in normal
     frames, and the blue line shows those in inverted frames.  The
     number of real emission lines can be estimated by the difference
     between the two.}
   \label{figure:snhist}
\end{figure*}

\begin{figure*}
   \begin{center}
      %\FigureFile(160mm,80mm){zhist.eps}
      \FigureFile(160mm,80mm){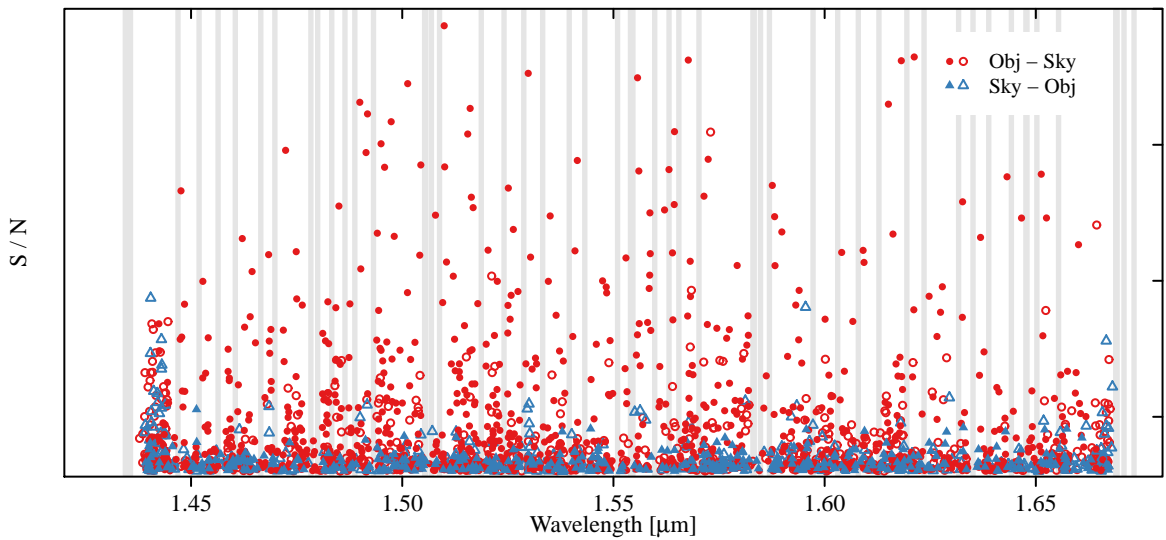}
   \end{center}
   \caption{The distribution of central wavelengths of the emission line
     candidates. The same symbols are used as Fig. \ref{figure:param}.
     Gray vertical stripes indicate the regions masked by the
     OH-airglow suppression mask.}
   \label{figure:zhist}
\end{figure*}

The cumulative $S/N$ distribution of emission line candidates
are displayed in Figure \ref{figure:snhist}.  The lower limit
to the number of real emission lines can be estimated from the
difference between the normal and inverted frames, i.e., red and blue
lines. (They are lower limits because the line detection
  completeness is not 100\% by statistical fluctuation and OH mask
  effects.  See the next paragraph.)  One can see that the number of
detections sharply increases with decreasing $S/N$ under $S/N=4.0$,
but this is mainly due to the increase of false detections in the
inverted frame.  The wavelength of the detected lines are plotted
against $S/N$ in Figure \ref{figure:zhist}.  Again, we see a tendency
for detections in the inverted images to be close to the OH mask
regions, indicating that residuals from the OH airglow is one of the
major reasons for false detections.

\begin{figure}
 \begin{center}
 \FigureFile(80mm,80mm){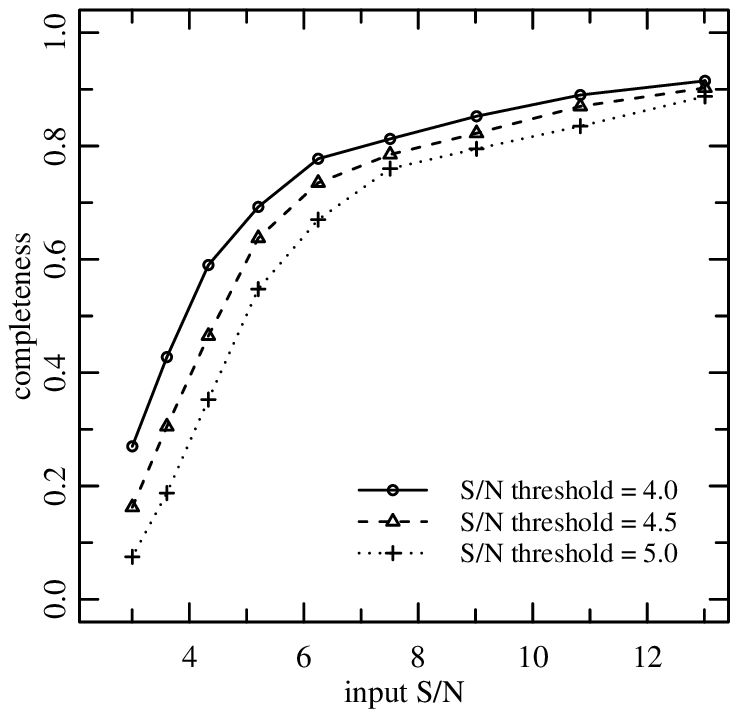}
 \end{center}
 \caption{The detection completeness as a function of input
     object brightness (in terms of $S/N$ computed with the noise
     level) for the three different values of the detection threshold
     $S/N$ indicated in the figure.}
 \label{figure:completeness}
\end{figure}

  The completeness of detection, i.e., the probability of
  successful detection for a given real emission line is also an
  important statistic to evaluate the performance of the software.  To
  estimate this we ran a simulation by placing artificial objects and
  then applying our detection algorithm. The completeness against
  input brightness ($S/N$) is displayed in Figure
  \ref{figure:completeness}, for some different values of detection
  $S/N$ thresholds.  The completeness does not reach 100\% but stays
  lower than 90\% at large input $S/N$, and this can be explained by
  the effect of OH masks that cover about 20\% of the observed
  wavelength range.  Indeed, we confirmed that the completeness
  becomes close to $100\%$ when we carried out the same simulation
  excluding the OH mask regions.  The completeness is about 40\% for
  the input $S/N$ same as the adopted $S/N$ threshold, which is a
  reasonable result expected from statistical fluctuation and the OH
  mask effect.
  
It should be noted that we assumed the same Gaussian PSF as the kernel
described in \S\ref{subsection:algorithm} for the artificial objects
distributed in the simulation.  Though the kernel is set to be similar
to typical line profiles, the completeness for real emission galaxies
is expected to be lower than this. A quantitative estimate of the
completeness including this effect would be model dependent about
the line profile distribution of a galaxy sample, which is beyond the
scope of this work.

%\begin{figure}
%   \begin{center}
%      \FigureFile(80mm,80mm){sn_obsha.eps}
%   \end{center}
%   \caption{The signal-to-noise ratio of emission line candidates
%   plotted against the measured line fluxes. The observed
%   fluxes are corrected for the effects of fiber aperture loss.}
%   \label{figure:snflux}
%\end{figure}

\section{Conclusion}\label{conclusion}

In this work we developed an automated emission line detection
software for FMOS on the Subaru Telescope, using the data from the
FastSound project, targeting H$\alpha$ emission line galaxies at
$z\sim1.3$.  Emission line detection is based on $S/N$ obtained by a
convolution between the two-dimensional science frame and a
line-profile kernel.  A unique feature of the software for the FMOS
data having many OH airglow suppression masks is the amplification of
noise level (and consequent suppression in the $S/N$ calculation) in
the masked region by use of a flat-field image. Bad pixels on the detectors
and pixels affected by cosmic-rays are removed from the $S/N$
calculation.  We also calculated six shape parameters for the detected
lines, and these are used for further rejection of false detections.
This is particularly useful to remove those spurious sources caused by unusual dark patterns when the
detectors are in unstable condition.

The efficiency and reliability of the line detection were examined by
applying the method to inverted science frames obtained by exchanging the
object and sky images in the reduction process.  The false-detection
rate is $0.3$, $1.8$ and $5.2\%$ for $S/N$ above $5.0$, $4.5$, and $4.0$,
respectively. The emission line flux corresponding to $S/N = 5$ is
about $\sim 1.0 \times 10^{-16}\;{\rm [erg/cm^2/s]}$ in normal
condition.  The software is open to the community, and currently
available on request to the authors.

This work was supported in part by JSPS KAKENHI Grant Number 23684007.

\end{document}